\documentclass[10pt,twocolumn,english,aps,prl,showpacs,superscriptaddress,groupaddress,floatfix,graphics,graphicx]{revtex4-2}

\usepackage[T1]{fontenc}
\usepackage{color}
\usepackage[english]{babel}
\usepackage{float}
\usepackage{amsmath}
\usepackage{amssymb}
\usepackage{graphicx}
\usepackage{xcolor}
\usepackage{soul}
\usepackage[bookmarks]{hyperref}
\hypersetup{colorlinks,linkcolor=blue,citecolor=blue,urlcolor=blue}
\usepackage[all]{hypcap}

\usepackage{soul}
\usepackage{comment}
\usepackage{orcidlink}

\usepackage{amsmath}
\usepackage{bm}
\usepackage{siunitx}
\usepackage{nicefrac}

\usepackage{graphicx}
\usepackage{xcolor}
\usepackage{xspace}
\usepackage{float}

\newcommand{\ket}[1]{| {#1} \rangle} 
\newcommand{\bra}[1]{\langle {#1} |} 
\DeclareMathAlphabet\mathbfcal{OMS}{cmsy}{b}{n}

\begin{document}
\preprint{AIP/123-QED}
\title{
Electrode-tunable nonlocal charge to spin conversion in WSe$_2$-intercalated bilayer graphene
}
\author{Marko Milivojevi\'c}
\email{marko.milivojevic@savba.sk}
\affiliation{Institute of Informatics, Slovak Academy of Sciences, 84507 Bratislava, Slovakia}
\affiliation {Faculty of Physics, University of Belgrade, 11001 Belgrade, Serbia}
\author{Juraj Mnich}
\email{juraj.mnich@student.upjs.sk}
\affiliation{Institute of Physics, Pavol Jozef \v{S}af\'{a}rik University in Ko\v{s}ice, 04001 Ko\v{s}ice, Slovakia}
\author{Martin Gmitra}
\affiliation{Institute of Physics, Pavol Jozef \v{S}af\'{a}rik University in Ko\v{s}ice, 04001 Ko\v{s}ice, Slovakia}
\affiliation{Institute of Experimental Physics, Slovak Academy of Sciences, 04001 Ko\v{s}ice, Slovakia}
\affiliation{New Technologies Research Centre, University of West Bohemia, Univerzitní 8, CZ-301 00 Pilsen, Czech Republic}
\begin{abstract}
We show that intercalating a WSe$_2$ monolayer into bilayer graphene mediates wavefunction hybridization of the two graphene layers at the Fermi level.
These delocalized states entangle both graphene layers, creating a synthetic bilayer graphene with interlayer coupling comparable to the proximity-induced spin-orbit interaction.
By analyzing transport properties of a four-terminal device, we demonstrate equal entangled parallel charge currents in both graphene layers, allowing us to unlock the hidden Rashba states via layer-selective chirality manifested in opposite-signed local Rashba-Edelstein signals.
We also show nonlocal Rashba-Edelstein and spin Hall effects activated in one graphene layer when a charge current is driven in a spatially adjacent graphene layer.
%
%
%
This effect is robust to the twist angle modulation between graphene and WSe$_2$, and to the applied electric field, suggesting its durability.
%


\end{abstract}
\maketitle
Van der Waals (vdW) heterostructures~\cite{Geim2013} provide a platform to fine-tune electronic properties via interlayer wavefunction overlap, enabling proximity-induced spin-orbit coupling (SOC)~\cite{GIK+17,FSY+23}, topological phases~\cite{ZF25}, and correlated states~\cite{ZKF24}, beyond what is possible in isolated monolayers. Thus, stacking of two-dimensional layers into vdW heterostructures can transform conventional materials into devices with novel spintronic effects~\cite{Sierra2021,ZLA25}.

For spintronics applications, graphene~\cite{NGM+04} has been recognized as a potentially interesting platform~\cite{ZFS04,FME+07} due to its long spin diffusion length~\cite{HGN+06,TJP+07} and high mobility~\cite{NGM+04}, that enable spin injection~\cite{OSN+07}, transport~\cite{NGM+04}, and detection~\cite{GOW18}, as demonstrated in devices like spin valves~\cite{CCF07,XSK+18} and field-effect transistors~\cite{ZFS04,FME+07,HKG+14}. For spin-based logic, it is crucial to achieve efficient spin manipulation~\cite{SKZ+17,LSC+19} and charge-spin conversion~\cite{Ghiasi2019,CSC+22,Lee2022b,OSH+23,Chi2024,CDY+25} in graphene without the need for large external magnetic fields, which are difficult to control at the nanoscale. The spin-orbit proximity provides this effect~\cite{Avsar2014,Gmitra2015}, using which it is possible to control the spin currents~\cite{YTL+16} in the target proximitized material.

The standard approach to induce sizable SOC in monolayer/bilayer graphene is to place it in a vdW heterostructure with materials that have strong SOC, typically transition-metal dichalcogenides~\cite{ZCS11,Gmitra2017,ZGF20,Zollner2021}. Besides transition metal dichalcogenides, stacking graphene with ferroelectric materials can lead to novel functionalities, such as the ferroelectric switching of the spin current~\cite{MMJ+26} and in-plane anisotropic spin relaxation rate~\cite{MGK+24}, with the latter experimentally confirmed~\cite{SST+25,SSC+26}. 

An established approach to transforming graphene into a spintronic material is based on two assumptions. Firstly, the horizontal mirror plane ($\sigma_{\rm h}$) must be broken, thereby triggering Rashba splitting and enabling sizable charge-spin conversion.
This is done by proximitizing graphene with a strong-SOC material~\cite{GKB+19,ZHK20,HSI+20,Naimer2021,HKZ+21,FMZ+21,IGH+22,KSM+23,YMK+24,MGP+24,GEM+25,RSC+26}. The second assumption is that proximity effects operate in a perturbative regime, in which the graphene Dirac points lie outside the interlayer hybridization window, so that proximity-induced spin-orbit coupling and interlayer tunneling act on well-separated energy scales.

In this Letter, we challenge both assumptions by studying graphene/WSe$_2$/graphene, a heterostructure in which WSe$_2$ acts not as a passive substrate but as an active mediator of interlayer wavefunction overlap, hybridizing the top and bottom graphene layers at the Fermi level and forming a {\it synthetic bilayer graphene}. 
Because the graphene Dirac points here lie within the interlayer hybridization window, proximity-induced SOC and interlayer tunneling act on comparable energy scales, placing the system beyond the perturbative regime of standard single-interface proximity heterostructures. 
As a consequence, a nonlocal charge-spin conversion is triggered, in which the charge current injected into one graphene layer generates charge current and spin accumulation detectable in the opposite layer. We also demonstrate that the proposed functionality persists when WSe$_2$ is twisted relative to the graphene layers, as well as under an applied electric field that introduces a slight asymmetry between the top and bottom graphene layers.

We first return to the untwisted, zero-field configuration, where the $\sigma_{\rm h}$ symmetry of the stack imposes compensation of local Rashba contributions from the top and bottom graphene/WSe$_2$ interfaces, giving rise to a {\it hidden Rashba effect}~\cite{YLZ+19} in which each graphene layer hosts an in-plane spin texture of opposite chirality despite the absence of an apparent Rashba splitting~\cite{ZLL+14, R60, BR84}. 
The opposite-sign charge-spin conversion signals in each layer represent an experimental fingerprint of the hidden Rashba physics. In contrast, the finite cross-layer signal represents direct evidence of the WSe$_2$-mediated interlayer hybridization as a new channel for nonlocal spin generation. The studied heterostructure thus represents an experimentally accessible platform for next-generation vdW spintronic devices that combines proximity-induced SOC, hidden Rashba selectivity, and cross-layer charge-spin conversion in synthetic bilayer stacks.

\begin{figure}[t]
    \centering
\includegraphics[width=0.99\linewidth]{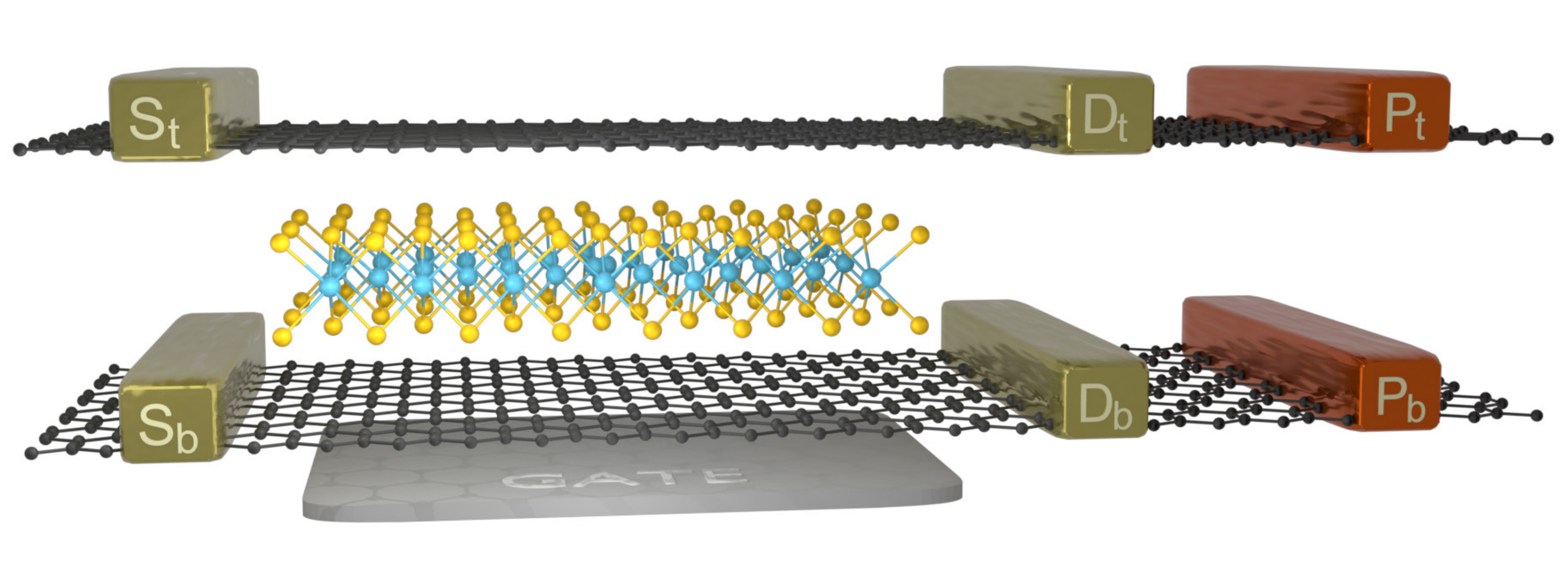}
    \caption{
{\bf Schematics of the WSe$_2$-intercalated bilayer graphene device}. 
The WSe$_2$ 
monolayer mediates interlayer overlap between the top and bottom graphene layers, forming a {\it synthetic bilayer graphene} at the Fermi level. 
Source (S), drain (D) and probe (P) electrodes on the top (S$_{\rm t}$, D$_{\rm t}$, P$_{\rm t}$) and bottom (S$_{\rm b}$, D$_{\rm b}$, P$_{\rm b}$) graphene layers enable three different transport setups: 
(1) all active electrodes probe the fully symmetric device and confirm the compensation of the Rashba effect; 
(2) lateral activation reveals the hidden Rashba effect in the top/bottom graphene layer; 
(3) crossed signal demonstrates hybridization-mediated nonlocal charge-spin conversion.
}
\label{fig:setup}
\end{figure}

In the studied graphene/WSe$_2$/graphene heterostructures with ${\bf D}_{3{\rm h}}$ symmetry, the top and bottom graphene differ only by a vertical shift, with zero twist angle between WSe$_2$ and the top/bottom graphene. 
More details on the heterostructure, computed via first-principles calculations with Quantum ESPRESSO~\cite{QE1,QE2} code, with the PBE functional~\cite{PBE}, scalar-relativistic SG15 ONCV pseudopotentials~\cite{H13,SG15,SGH+16} (fully relativistic for SOC calculations), Grimme-D2 vdW corrections~\cite{G06,BCF+08}, 1 mRy Methfessel-Paxton smearing~\cite{MP89}, $6\times6$ $k$-mesh, 70 Ry wavefunction/280 Ry charge density cutoffs, dipole corrections~\cite{B99}, and 20~\AA\ vacuum, is provided in the Supplementary Material (SM)~\cite{SUPP}. Structures were relaxed using the quasi-Newton scheme to force and energy thresholds of $10^{-4}$ Ry/bohr and $10^{-7}$ Ry, respectively.

\begin{figure}[t]
    \centering
\includegraphics[width=0.9\linewidth]{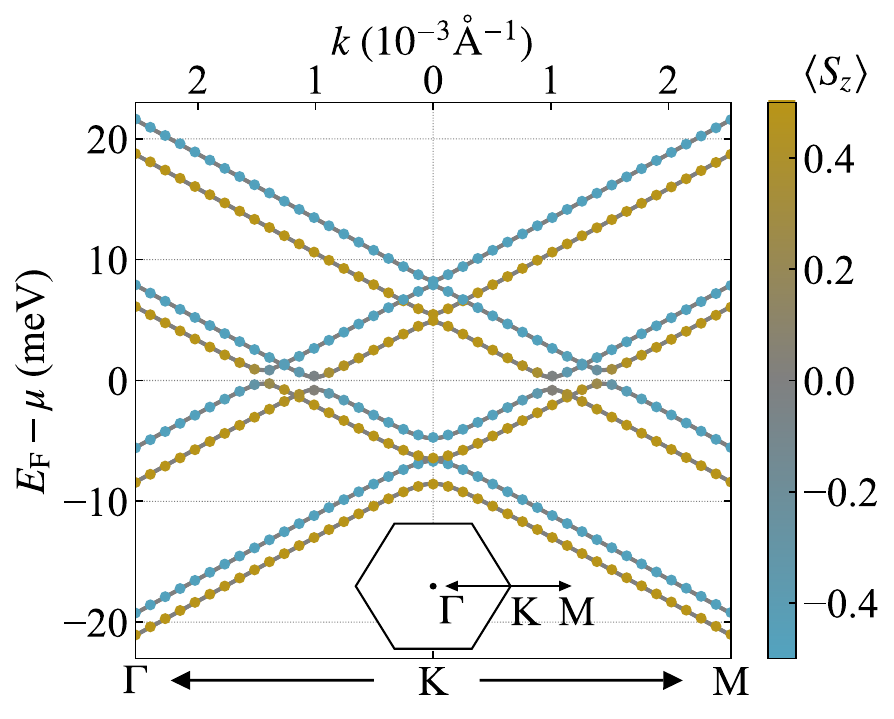}
    \caption{First-principles electronic structure calculations of the zero twist angle graphene/WSe$_2$/graphene heterostructure in the vicinity of the graphene K point for zero doping, $\mu=0$.
    The color map corresponds to the $z$-component of the spin expectation value.
    The solid lines are dispersions calculated using the effective model.
    }\label{fig:unfold}
\end{figure}

In FIG.~\ref{fig:unfold} we plot the band structure in the vicinity of the Fermi level along the 
$\Gamma$KM path of graphene's Brillouin zone.
The electronic structure is a combination of two Dirac cones with an offset of 13~meV and prominent proximity-induced spin splitting.
The hybridization gap of 0.6~meV opens at the Fermi level as a result of WSe$_2$ mediated hybridization between the Dirac electrons of the two adjacent graphene layers.
We note that the band topology lies deep within the semiconducting gap of WSe$_2$.
Atomic wave projections confirm that the bands near the K point are of pure graphene character, with equal contributions from the top and bottom layers, consistent with $\sigma_{\rm h}$ symmetry and further supported by atom-projected density of states (see Fig.~S1 of SM~\cite{SUPP}). 
These projections reveal that the mediated hybridization across the WSe$_2$ monolayer forms the {\it synthetic bilayer graphene} with layer-delocalized entangled states. 
Equal contributions of both graphene layers result in the absence of in-plane spin expectation values, confirming a compensated Rashba effect

{\it Model for synthetic bilayer graphene}. 
To model the synthetic graphene bilayer at the Fermi level, we first revisit the effective Hamiltonian of proximitized single-layer graphene, and then introduce WSe$_2$-mediated tunneling terms capturing the interlayer interaction between graphene layers. For a single-layer graphene with ${\bf C}_{3{\rm v}}$ symmetry~\cite{Kochan2017}, the effective Hamiltonian can be written as $H_{\rm gr}=H_0+H_{\rm I}+H_{\rm{R}}$. Here $H_0$ represents the orbital Hamiltonian, whose first part is the sublattice-dependent on-site potential, $\sum_{i=\rm{A},\rm{B}}\sum_{\sigma}(\mu+\Delta (-1)^{\delta_{\rm{B},i}})c_{i\sigma}^{\dagger}c_{i\sigma}$,
equal to $\mu\pm \Delta$ on sublattice A/B of graphene, where $\mu$ is the chemical potential, $\Delta$ is the stagerred potential, and $c_{i\sigma}^{\dagger}$ ($c_{i\sigma}$) is the creation (anihilation) operator on site $i=\rm{A},\rm{B}$ with spin $\sigma=\pm$. 
Furthermore, the orbital Hamiltonian consists of a nearest-neighbor Hamiltonian with hopping $t$, $-t \sum_{{\langle m,n\rangle}}\sum_{\sigma}c_{n\sigma}^{\dagger}c_{m\sigma}$, which can be connected to the Fermi velocity $v_{\rm F}$ through the relation $v_{\rm F}=a t \sqrt{3}/2\hbar$. 
The second term $H_{\rm I}=\sum_{\xi={\rm A,B}}\sum_{\langle\langle m,n\rangle\rangle}\frac{{\rm i}\alpha_{\rm I}^{\xi}}{3\sqrt{3}}\sum_{\sigma}\nu_{m,n}[s_z]_{\sigma\sigma}c_{m\sigma}^{\dagger}c_{n \sigma}$ describes the intrinsic SOC, where the sum goes over the next-nearest-neighbors interaction described in terms of the sublattice-dependent $\alpha_{\rm I}^{\rm{A}/\rm{B}}$ parameters and the sign factor $\nu_{m,n}$ that has the value 1 ($-1$) when the next-nearest-neighbor hopping from site $m$ to site $n$ 
via the common nearest-neighbor encloses a clockwise (counterclockwise) path.  The last term, $H_{\rm{R}}$, describes the nearest-neighbor Rashba SOC term, 
$H_{\rm{R}}=\frac{2{\rm i}\alpha_{\rm R}}{3}\sum_{\langle m,n\rangle}\sum_{\sigma\neq\sigma'}
[{\bf s}\times {\bf d}_{m,n}]_{\sigma\sigma'}^z c_{m\sigma}^{\dagger}c_{n\sigma'}$, 
where ${\bf s}$ is the vector of Pauli matrices, $\alpha_{\rm R}$ represents the Rashba SOC strength, ${\bf d}_{m,n}$ is the unit vector in the horizontal plane pointing from lattice site $n$ to the nearest-neighbor site $m$.

In the studied heterostructures, the effective parameters of the top and bottom graphene layers are connected by horizontal mirror symmetry $\sigma_{\rm h}$. The horizontal mirror symmetry transforms the top/bottom graphene atom on sublattice A/B to the bottom/top atom on sublattice A/B. A simple group-theoretical analysis reveals that the spin-independent and spin-conserving parameters of the top and bottom graphene, $v_{\rm F},\mu,\Delta,\alpha_{\rm{I}}^{\rm A},\alpha_{\rm{I}}^{\rm B}$, remain unchanged. In contrast, the Rashba parameter $\alpha_{\rm R}$ acquires a sign change, $\alpha_{\rm R}^{\rm t}=-\alpha_{\rm R}^{\rm b}$. 

The intercalated WSe$_2$ monolayer mediates the interaction between the graphene monolayers, as evidenced by the projected-wave-function analysis.
We propose the spin-conserving valley-dependent effective interlayer interaction Hamiltonian valid in the vicinity of the $\kappa {\bm K}$ point, 
\begin{equation}
    H_{\rm int}({\bf k})= \frac{1}{2}\sum_{i={\rm{A,B}}} 
    (\sigma_0+(-1)^{\delta_{{\rm{B}},i}}\sigma_z)\otimes(t_i s_0+\kappa\tau_i s_z), 
\end{equation}
where $\kappa=\pm1$ is the valley index, $\sigma_0/s_0$ and $\sigma_z/s_z$ are Pauli matrices in the pseudospin/spin space, $t_{\rm A}$ and $t_{\rm B}$ represent spin-independent tunneling amplitudes determining the Dirac cones offset, whereas $\tau_{\rm A}/\tau_{\rm B}$ correspond to Ising-type spin-dependent tunneling coefficients affecting asymmetry of the avoided crosings at the Fermi level. 
The effective model of the synthetic bilayer graphene $ H_{\rm sbg}({\bf k})$ in the vicinity of the $\kappa {\bm K}$ point reads
\begin{equation}\label{BGReff}
    H_{\rm sbg}({\bf k})=\begin{pmatrix}
H_{\rm gr}^{\rm t}({\bf k}) & H_{\rm int}({\bf k}) \\
H_{\rm int}^{\dagger}({\bf k}) & H_{\rm gr}^{\rm b} ({\bf k})
\end{pmatrix},
\end{equation}
where the Hamiltonian $H_{\rm gr}^{\rm t/b}({\bf k})$ describes the effective Hamiltonian of the top/bottom graphene,
whereas $ H_{\rm int}({\bf k})$ describes effective tunneling between the graphene layers. 
\begin{table}[t]
\caption{Parameters of the effective synthetic bilayer graphene model for the graphene/WSe$_2$/graphene heterostructure.}
\label{TAB:parameters}
\centering
\small
\setlength{\tabcolsep}{4.6pt}
\renewcommand{\arraystretch}{1.0}
\begin{tabular}{lr|cc}
\hline
\multicolumn{2}{c|}{top/bottom graphene} & \multicolumn{2}{c}{interacting terms} \\\hline\hline
$v_{\rm F}^{\rm t/b}$ [$10^{6}${\rm m/s}] & 0.813&&\\
$\Delta^{\rm t/b}$\;[{\rm meV}]        &$-0.343$&$t_{\rm A}$\;[{\rm meV}] & 7.199\\
$\mu^{\rm t/b}$\;[{\rm meV}]            &$-1.142$&$t_{\rm B}$\;[{\rm meV}] & 5.939\\
$\alpha_{\rm R}^{\rm t/b}$\;[{\rm meV}] & $-0.540/0.540$&$\tau_{\rm A}$\;[{\rm meV}] & -0.171\\ 
$(\alpha_{\rm I}^{\rm A})^{\rm t/b}$\;[{\rm meV}] &  1.170&$\tau_{\rm B}$\;[{\rm meV}] & -0.358\\
$(\alpha_{\rm I}^{\rm B})^{\rm t/b}$\;[{\rm meV}] & $-1.174$&&\\\hline
\end{tabular}
\end{table} 

Using this model, we were able to fit the DFT data, see Fig.~\ref{fig:unfold}. 
The results, given in Table~\ref{TAB:parameters}, reveal significant interaction between the graphene layers, characterized by the dominant energy scale of the spin-independent hopping parameters $t_{\rm{A}}$ ($\sim7.2$\,meV) and $t_{\rm{B}}$ ($\sim6.0$\,meV). 
In addition, nonzero Rashba parameters ($\alpha_{\rm R}^{\rm t}=-\alpha_{\rm R}^{\rm b}=-0.54$\,meV) indicate hidden Rashba effects in top and bottom graphene with opposite chirality. Compared to bilayer graphene~\cite{CK13}, this interlayer interaction is two orders of magnitude smaller and matches the scale of the proximity-induced SOC \cite{Gmitra2017}. 
This provides a unique platform that combines interlayer entanglement and proximity effects, which must be treated on equal footing.
Additionally, realization of the synthetic bilayer is robust against twisting and gating effects.
The former can be modeled by a modified $H_{\rm gr}^{\rm t/b}({\bf k})$ Hamiltonian to incorporate the presence of the Rashba phase ${\phi_{\rm R}}$, triggered by the nonzero twist angle~\cite{David2019, Li2019b}.
In the gating case, the asymmetry between top and bottom graphene primarily affects the chemical potentials, leading to $\mu^{\rm b}\neq\mu^{\rm t}$;
for further details, see the SM~\cite{SUPP}. 

{\it Nonlocal spin transport.} 
The emergence of interlayer hybridization enables nonlocal spin transport, which can be detected in standard charge-spin conversion experiments. The presence of the Rashba SOC within each graphene layer, i.e., the in-plane spin texture, is responsible for the appearance of the Rashba-Edelstein effect (REE)~\cite{E90,DBD14,OMR+17}. Assuming that the charge current, $\delta J_x$, is applied in the $x$-direction (direction of the applied bias voltage $V_{\rm{b}}$), the Rashba-Edelstein conversion efficiency, $\alpha_{\rm REE}$, can be defined as $\alpha_{\rm REE}=e v_{\rm{F}}/\hbar ({\delta S_y}/{\delta J_x})$, where $\delta S_y$ is the current-induced nonequilibrium spin density along the $y$-axis. We calculate the Rashba-Edelstein coefficient $\alpha_{\rm REE}$ using Kubo formula~\cite{K56,K57} in the Smrčka-Středa formulation~\cite{SS77,CB01,BM20}. Assuming the weak disorder scattering described by the phenomenological parameter $\gamma$~\cite{Lee2022b,BM20,FBM14,ZZF+17}, we defined the change $\delta \mathcal{O}$ of the physical quantity represented by the operator $\mathcal{O}$ as $\delta  \mathcal{O}=\frac{E}{4\pi^2}\int d^2{\bf k}[\chi_{\mathcal{O}}^{\rm surf}({\bf k})+\chi_{\mathcal{O}}^{\rm sea}({\bf k})]$, where $\mathcal{O}\in\{S_{y},J_{x}\}$, $S_{y}=\frac{\hbar}2 s_{y}$, and $J_{x} = -ev_{x}$. The Fermi surface and Fermi sea susceptibilities, $\chi_{\mathcal{O}}^{\rm surf}$ and $\chi_{\mathcal{O}}^{\rm sea}$~\cite{BM20}, are equal to $\chi_{\mathcal{O}}^{\rm surf}({\bf k})=\frac{\hbar}{\pi}\gamma^2\sum_{n,m} {\rm Re}\{
[\mathcal{O}J_x]_{nm}\}/\varepsilon_{nm}^{{\bf k}\gamma}$ and       $\chi_{\mathcal{O}}^{\rm sea}({\bf k})=\hbar
    \sum_{n,m \neq n} (f_{n,{\bf k}}-f_{m,{\bf k}})
    {\rm Im}\{
[\mathcal{O}J_x]_{nm}\}/
    (\epsilon_{n{\bf k}}-\epsilon_{m{\bf k}})^2$, 
where $[\mathcal{O}J_x]_{nm}=\bra{n{\bf k}}\mathcal{O}\ket{m{\bf k}} \bra{m{\bf k}}J_x\ket{n{\bf k}}$, 
$\varepsilon_{nm}^{{\bf k}\gamma}=[(\epsilon_{n{\bf k}}-E_{\rm F})^2+\gamma^2][(\epsilon_{m{\bf k}}-E_{\rm F})^2+\gamma^2]$, 
whereas $\epsilon_{n{\bf k}}$ and $\ket{n {\bf k}}$, $n=1,\dots,8$, represent eigenvalues and eigenvectors of the $H_{\rm sbg}$ in the vicinity of the $K/K'$ points, $f_{n,{\bf k}}$ is the Fermi function, and $\gamma$ is the quasiparticle lifetime. In all calculations of conversion coefficients, we performed the integration of $\chi_{\mathcal{O}}^{\rm sea}$ and $\chi_{\mathcal{O}}^{\rm surf}$ around the $K$ and $K'$ points.
This integration was carried out on a square grid, centered around K(K'), having the size of $0.008$~\AA$^{-1}$ and a step $\Delta k=5 \times 10^{-7}$~\AA$^{-1}$. All calculations were performed at an electronic temperature $k_{\rm B}T=0.01$~meV.

We present three experimental setups to reveal distinct charge-spin conversion features of the synthetic bilayer graphene device sketched in FIG.~\ref{fig:setup}.
To this end, we configure source, drain, and probe electrodes on the top (S$_{\rm t}$, D$_{\rm t}$, P$_{\rm t}$) and bottom (S$_{\rm b}$, D$_{\rm b}$, P$_{\rm b}$) graphene layers, respectively. 
These are selectively activated by passing current between source and drain and measuring the response via probe electrodes, usually in nonlocal spin experiments with one ferromagnetic electrode.
We distinguish three distinct transport situations: 
\emph{setup~1}, where the charge current flows through both graphene layers with all source and drain electrodes active, and probe electrodes detect the overall signal; 
\emph{setup~2} (lateral configuration),
where both the charge current flow and the resulting spin accumulation are measured within the same selected layer; and 
\emph{setup~3} (crossed configuration), where the charge current flows in one graphene layer, and the response is measured in the adjacent graphene layer.
To address the nonequilibrium physical quantitities to individual graphene layers for the considered setups, the appropriate operators $\mathcal{O}$ were projected onto top or bottom layer $\mathcal{O}^{\rm t/b}=(\mathcal{P}^{\rm t/b})^{\dagger}\mathcal{O}\mathcal{P}^{\rm t/b}$, using the projection operators $\mathcal{P}^{\rm t}={\rm diag}[1,1,1,1,0,0,0,0]$ and $\mathcal{P}^{\rm b}={\rm diag}[0,0,0,0,1,1,1,1]$. 
Using this definition, the REE efficiency coefficient in the case of charge current flow between S$_{\rm t}$ and D$_{\rm t}$ electrodes, and response measured with P$_{\rm t}$ probe, is equal to 
$\alpha_{\rm REE}^\text{tt}=e v_{\rm{F}}/\hbar (\delta S_y^{\rm t}/{\delta J_x^{\rm t}})$.
Similarly, we define the other coefficient for the lateral configuration
$\alpha_{\rm REE}^\text{bb}=e v_{\rm{F}}/\hbar (\delta S_y^{\rm b}/{\delta J_x^{\rm b}})$, and the coefficients for the crossed configurations
$\alpha_{\rm REE}^\text{tb}=e v_{\rm{F}}/\hbar (\delta S_y^{\rm b}/{\delta J_x^{\rm t}})$, and 
$\alpha_{\rm REE}^\text{bt}=e v_{\rm{F}}/\hbar (\delta S_y^{\rm t}/{\delta J_x^{\rm b}})$.
\begin{figure}[h]
    \centering
\includegraphics[width=0.9\linewidth]{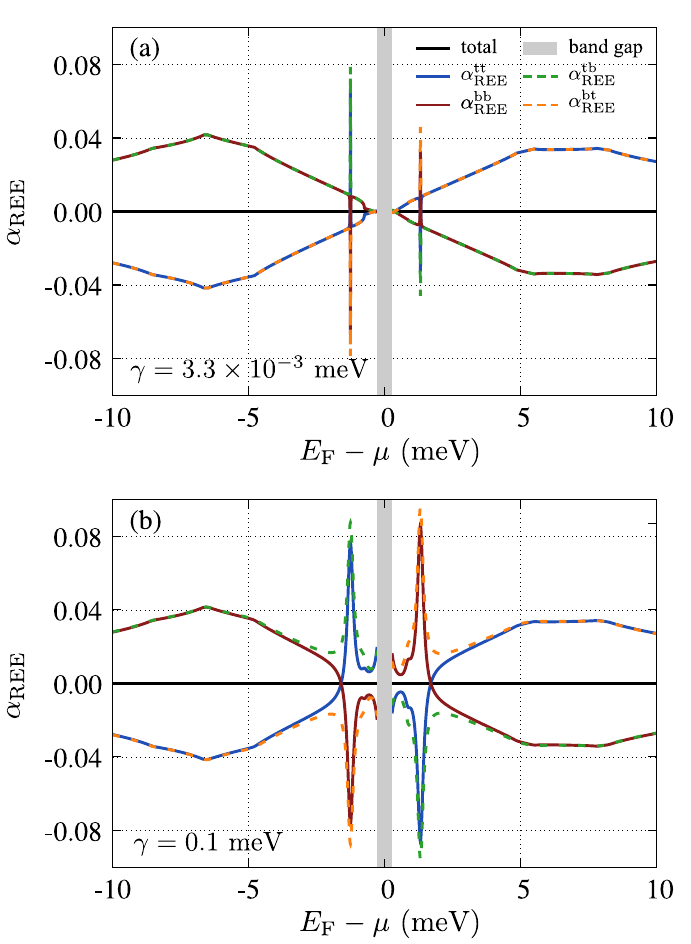}
    \caption{Calculated charge-spin conversion coefficients $\alpha_{\rm REE}$ as a function of doping for three transport setups: (i)~setup~1 (total), with all electrodes active; (ii)~setup~2 (lateral configurations, $\alpha_{\rm REE}^\text{tt}$, $\alpha_{\rm REE}^\text{bb}$), where current drive and voltage probing occur in the same graphene layer; and (iii)~setup~3 (crossed configurations, $\alpha_{\rm REE}^\text{tb}$, $\alpha_{\rm REE}^\text{bt}$), where current is driven in one layer and probed in the adjacent layer. Results are shown for quasiparticle broadenings of 
    (a)~$\gamma=3.3\times10^{-3}$~meV and (b)~$\gamma=0.1$~meV.
    }
    \label{fig:cscFINAL}
\end{figure}

In FIG.~\ref{fig:cscFINAL} we present calculated charge-spin conversion coefficients $\alpha_{\rm REE}$ as a function of doping. 
In setup~1, $\alpha_{\rm REE}$ is zero, independently of doping, confirming compensation of the Rashba effect due to preservation of the global horizontal mirror plane symmetry $\sigma_{\rm h}$. 
We note that in real devices, this symmetric configuration can be achieved with a dual-gated geometry, which enables independent control of both carrier density and displacement fields.
In setup~2, the lateral transport configurations, the nonzero oppositelly signed $\alpha_{\rm REE}^\text{tt}$ and $\alpha_{\rm REE}^\text{bb}$ are observed
for the same doping energy, demonstrating the layer-selective chirality switching due to the presence of the hidden Rashba effect in top and bottom graphene.
In setup~3, the crossed configuration, the calculated nonzero $\alpha_{\rm REE}^\text{tb}$ and $\alpha_{\rm REE}^\text{bt}$ coefficients confirm the generation of the hybridization-induced nonlocal spin polarization.
To assess the robustness of the nonlocal spin polarization, we investigate the effect of a finite quasiparticle lifetime. 
Sharp peaks in the spectrum at $E_{\rm{F}} - \mu \approx -1.25$~meV and $E_{\rm{F}} - \mu \approx +1.32$~meV stem from band crossings away from the $K$ point. 
As shown in Fig.~\ref{fig:cscFINAL}(a) for a small broadening of $\gamma=3.3\times10^{-3}$~meV, the lateral and crossed transport setups yield identical responses. 
For broadening on the order of the hybridization gap ($\gamma=0.1$~meV) [Fig.~\ref{fig:cscFINAL}(b)], the lateral transport setups reverse signs for doping between the sharp peaks; importantly, however, the crossed nonlocal signals maintain their sign as well as magnitude. 
Furthermore, this nonlocal charge-spin conversion persists under a finite electric field and for nonzero twist angles, confirming its robustness against gate-induced inversion asymmetry and twist variations [Sec.~3 of SM~\cite{SUPP}].

\begin{figure}[h]
    \centering
    \includegraphics[width=0.98\linewidth]{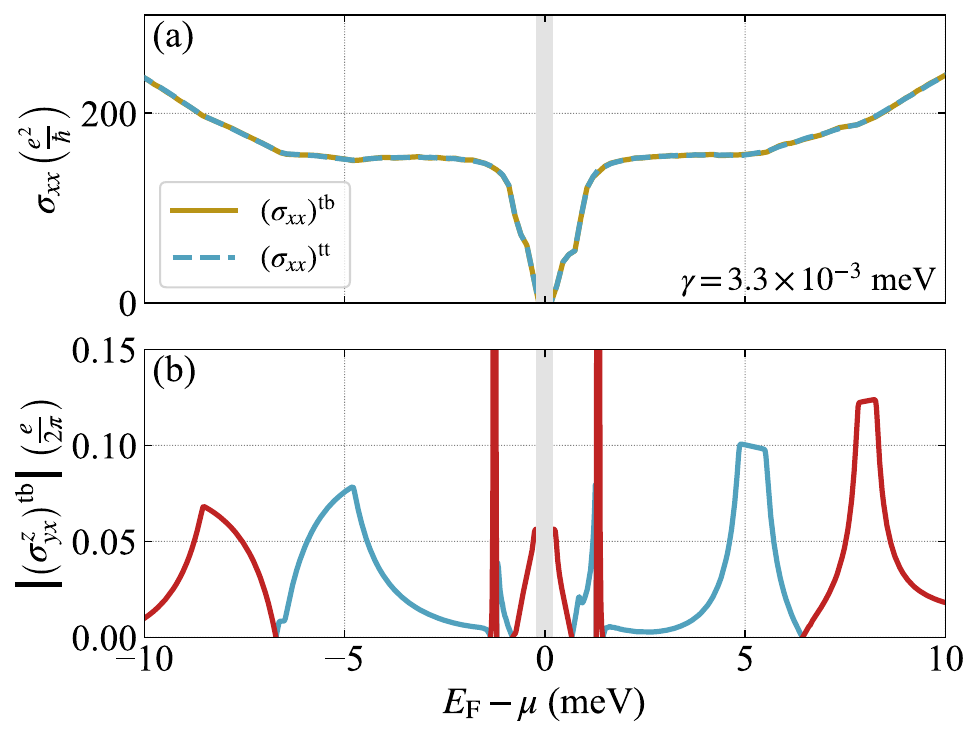}
    \caption{Calculated transport properties as a function of doping.
    (a)~Charge current in the top layer driven between S$_{\rm t}$ and D$_{\rm t}$ electrodes, and between the S$_{\rm t}$ and D$_{\rm b}$ electrodes.
    (b)~Absolute value of the nonlocal spin Hall conductivity $(\sigma_{yx}^z)^{\rm tb}$ probed in the bottom layer when driving charge current in the top graphene layer. Red branches are for positive and blue for negative values.
    The gray area is the bandgap.
    }
    \label{fig:SHE}
\end{figure}

While horizontal mirror symmetry completely suppresses the total REE effect, it has no impact on the unconditional existence of the spin Hall effect (SHE). 
The spin Hall current can be subsequently distinguished from the spin accumulation, e.g., the response to a magnetic field along the $z$-direction affects REE but not the SHE.
Hybridization of the electronic states in the synthetic bilayer graphene yields equal charge current in both layers irrespective of the driving bias, see Fig.~\ref{fig:SHE}(a).
To distinguish the pure nonlocal spin effects from the charge-driven contribution, we calculated the nonlocal SHE as the anticommutator $\mathcal{O} = [S_{z},v_{y}]_{+}$, in the bottom layer as a response to the charge current driven in the top layer only.
In Fig.~\ref{fig:SHE}(b), we show nonlocal spin Hall conductivity $(\sigma_{yx}^z)^{\rm tb}$ for the spin current along the $y$-direction, polarized in the $z$-direction in the bottom graphene layer when the charge current is driven in the top graphene layer.
The nonlocal-driven spin Hall conductivity $(\sigma_{yx}^z)^{\rm tb}$ exhibits antisymmetric-like behavior about the two Dirac cones at $-7$~meV and $6$~meV, and a peaked symmetric-like behavior around the Fermi level. 
The peaked structure reflects the band crossing and band edges of the hybridization gap, which has reversed sign for the lateral configurations $(\sigma_{yx}^z)^{\rm tt}$ and $(\sigma_{yx}^z)^{\rm bb}$.


{\it Conclusions.}
We demonstrated that the graphene/WSe$_2$/graphene heterostructure realizes a {\it synthetic bilayer graphene} platform, in which WSe$_2$ acts as an active mediator of interlayer wavefunction overlap, hybridizing the top and bottom graphene layers at the Fermi level. 
This interlayer entanglement operates at the proximity-induced SOC energy scale.
By investigating transport properties, we addressed three spintronic signatures:
(i)~a vanishing Rashba-Edelstein signal confirming the absence of Rashba SOC enforced by the global $\sigma_{\rm h}$ symmetry; 
(ii)~opposite chiralities from top and bottom layers, evidencing a {\it hidden Rashba effect}; and 
(iii)~finite cross-layer signals, demonstrating nonlocal charge-spin conversion and spin Hall signals. 
The presence of the nonlocal effects remains robust across different twist angles and perpendicular electric fields.
Intercalating bilayer systems establish a framework for investigation of a broader class of devices with cross-layer charge-spin conversion, paving the way for nonlocal spin generation and gate-tunable spintronics.

{\it Acknowledgments.}
Research results were obtained using the computational resources procured in the national project National competence centre for high performance computing (project code: 311070AKF2) funded by the European Regional Development Fund, EU Structural Funds Informatization of society, Operational Program Integrated Infrastructure.
M.M. acknowledges the financial support by the EU NextGenerationEU through the Recovery and Resilience Plan for Slovakia under the Project No. 09I02-03-V01-00012, by the APVV grant APVV-23-0430, and VEGA grants 2/0081/26 and 2/0133/25.
M.G.~acknowledges funding by the EU NextGenerationEU through the Recovery and Resilience Plan for Slovakia under the project No. 09I05-03-V02-00071, the Ministry of Education, Research, Development and Youth of the Slovak Republic, provided under Grant No. VEGA 1/0104/25, the Slovak Research and Development Agency under the Contract no APVV-25-0066, the Slovak Academy of Sciences project IMPULZ IM-2021-42, and support of the QM4ST project funded by Programme Johannes Amos Commenius, call Excellent Research (Project No. CZ.02.01.01/00/22\_008/0004572).\\
{\it Data Availability.} The data that support the findings of this study will be made publicly available on Zenodo upon acceptance of the manuscript.\\
{\it Author Contributions.} M.M. conceived and led the project, performed all first-principles and tight-binding calculations, developed the charge-spin conversion code, analyzed all results, and wrote an initial version of the manuscript. 
J.M. performed original charge-spin conversion calculations, spin Hall calculations, and conceived the experimental transport realization.
M.G. framed the core transport concept idea and prepared the final version of the manuscript.
\section*{References}
\bibliography{biblio}
\end{document}


\date{}
\maketitle

\noindent
\textsuperscript{1}Institute of Informatics, Slovak Academy of Sciences, 84507 Bratislava, Slovakia\\
\textsuperscript{2}Faculty of Physics, University of Belgrade, 11001 Belgrade, Serbia\\
\textsuperscript{3}Institute of Physics, Pavol Jozef \v{S}af\'{a}rik University in Ko\v{s}ice, 04001 Ko\v{s}ice, Slovakia\\
\textsuperscript{4}Institute of Experimental Physics, Slovak Academy of Sciences, 04001 Ko\v{s}ice, Slovakia\\
\textsuperscript{5}New Technologies Research Centre, University of West Bohemia, Univerzitní 8, CZ-301 00 Pilsen, Czech Republic\\
\textsuperscript{*}Email: marko.milivojevic@savba.sk\\
\textsuperscript{$\dagger$}Email: juraj.mnich@student.upjs.sk

\section{Details of the DFT calculation}
The lattice parameter of graphene is taken as $a_0=2.46~{\rm\AA}$, while for WSe$_2$ the latice parameter is equal to 
$a_{\rm WSe_2}=3.31~{\rm\AA}$~\cite{ZCS11}. The supercells were constructed by straining graphene, described by the relative strain $\delta$, using which the strained lattice parameter of graphene $a_0^{\rm str}$ can be described as $a_0^{\rm str}=(1+\delta[\%]/100)a_0$.  
Using the strained lattice vectors of graphene ${\bm a}_1=a_0^{\rm str}{\bm e}_x$ and ${\bm a}_2=a_0^{\rm str}(\cos{(2\pi/3)}{\bm e}_x+\sin{(2\pi/3)}{\bm e}_y)$ we can define lattice vectors of the heterostructure $(n_1,n_2)$ and $(m_1,m_2)$ as $(n_1,n_2)=n_1{\bm a}_1+n_2{\bm a}_2$ and $(m_1,m_2)=m_1{\bm a}_1+m_2{\bm a}_2$. In Table~\ref{TAB:structural} we list the twist angle between graphene top/bottom layer and WSe$_2$ used to create the heterostructure, considered heterostructure vectors $(n_1,n_2)$ and $(m_1,m_2)$, strain $\delta$ applied to graphene, the total number of atoms $N$ in the heterostructure, and the number of carbon atoms on the top and bottom layer of graphene. 

\begin{table}[t]
\caption{The structural information of the studied graphene/WSe$_2$/graphene heterostructures is given. Besides the relative twist angle $\theta$ between graphene monolayers and WSe$_2$, heterostructure lattice vectors $(n_1,n_2)$ and  $(m_1,m_2)$ are given in terms of the untwisted (but strained) lattice vectors of the graphene lattice: $(n_1,n_2)=n_1{\bm a}_1+n_2{\bm a}_2$; $(m_1,m_2)=m_1{\bm a}_1+m_2{\bm a}_2$. The
applied strain $\delta_{\rm Gr}$ to the graphene lattice parameter is also given, the total number of atoms $N$ in the heterostructure, and the number of carbon atoms on the top and bottom layer of graphene. Unstrained lattice constants of WSe$_2$ is 3.31~\cite{ZCS11}; in all calculated heterostructures WS$_2$ monolayer was twised, but not strained.}\label{TAB:structural}
\centering
\footnotesize
\setlength{\tabcolsep}{7pt}
\renewcommand{\arraystretch}{1.0}
\begin{tabular}{cccccc}
\hline\hline
$\theta\,[{\rm deg}]$  & $(n_1,n_2)$  & $(m_1,m_2)$ &  $\delta_{\rm Gr}$ [\%]& $N$ & $N_{\rm gr}^{\rm t/b}$\\\hline
0$^{\circ}$    & $(4,0)$  & $(0,4)$  & 0.915 & 91  & 32/32  \\\hline
5.2$^{\circ}$  & $(3,-1)$ & $(1,4)$  &-1.265 & 73  & 26/26  \\\hline 
19.1$^{\circ}$ & $(3,1)$  & $(-1,2)$ & 1.712 & 40  & 14/14  \\\hline 
27.0$^{\circ}$ & $(3,-1)$ & $(1,4)$  &-1.265 & 73  & 26/26  \\\hline 
\end{tabular}
\end{table}

We perform the electronic structure calculation of the graphene/WSe$_2$/graphene heterostructures using Density Functional Theory (DFT) as implemented in the plane wave code Q{\sc{uantum}} ESPRESSO (QE)~\cite{QE1,QE2}. The relaxation of the studied heterostructures was performed using the Perdew-Burke-Ernzerhof functional~\cite{PBE} and scalar-relativistic SG15 optimized norm-conserving Vanderbilt (ONCV) pseudopotentials~\cite{H13,SG15,SGH+16}. The kinetic energy cut-offs for the wave function and charge density were chosen to be 70~Ry and 280~Ry, respectively. Additionally, Methfessel–Paxton energy level smearing~\cite{MP89} of 1~mRy was used, and $6\times 6$ $k$-points mesh for the irreducible part of the Brillouin zone sampling were used for self-consistent calculations. 
The van der Waals interaction was modeled using the semiempirical Grimme-D2 correction~\cite{G06,BCF+08}, and a vacuum of 20~\AA~in the $z$-direction to detach the periodic images of the heterostructure was used. The positions of atoms were relaxed using the quasi-Newton scheme using scalar-relativistic pseudopotentials, keeping the force and energy convergence thresholds for ionic minimization to $1\times10^{-4}$~Ry/bohr and $10^{-7}$~Ry, respectively. 

For the self-consistent calculation, including the spin-orbit coupling (with and without the apllied perpendicular electric field), we use fully-relativistic ONCV pseudopotentials. Also, we have kept
the same $k$-mesh but increasing the energy convergence thresholds to $10^{-8}$~Ry. Additionally, dipole correction~\cite{B99} was applied to properly determine the Dirac point energy offset due to dipole electric field effects between graphene's and WSe$_2$.

\begin{figure}[h]
    \centering
\includegraphics[width=0.995\linewidth]{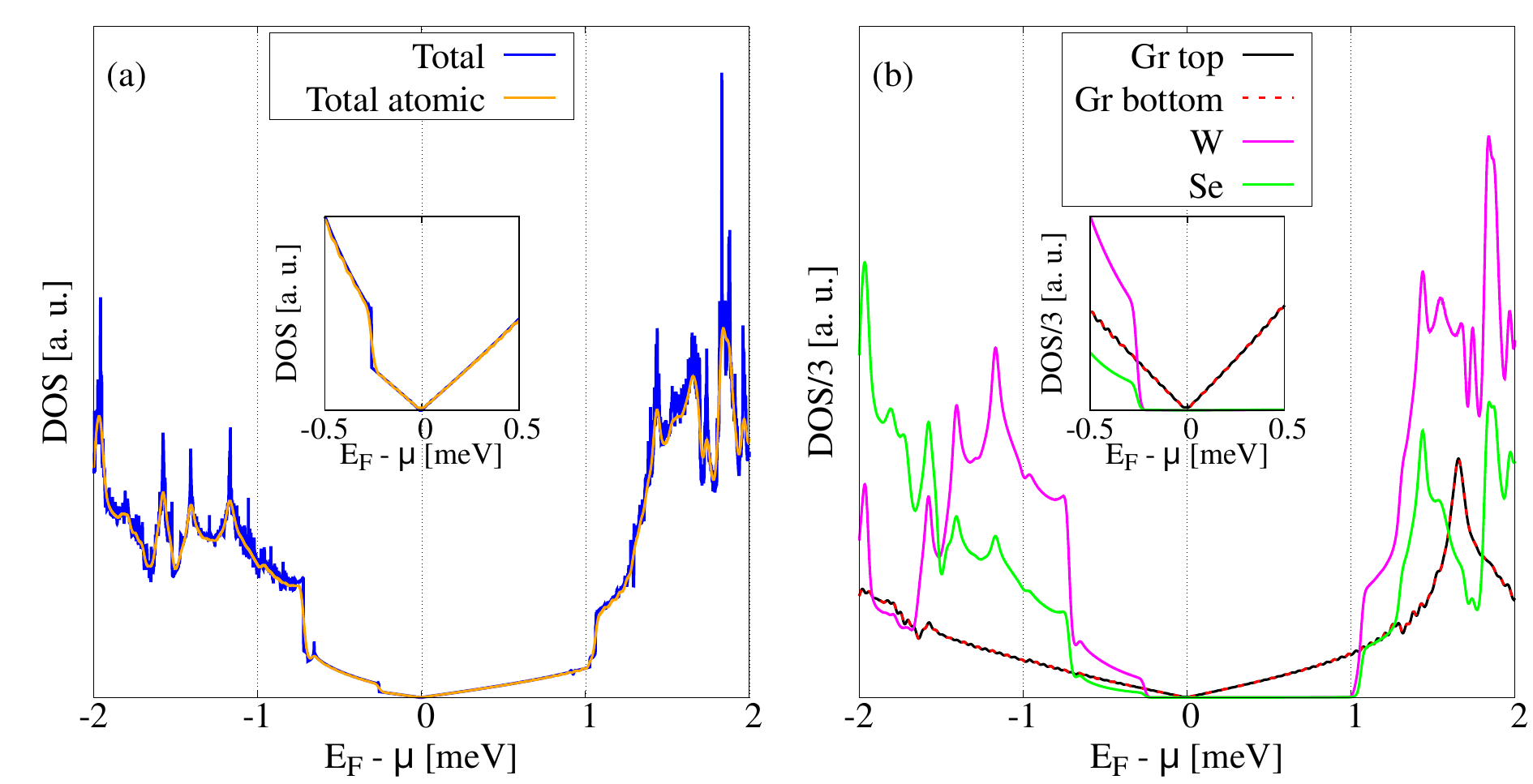}
    \caption{Density of states (DOS) of the graphene/WSe$_2$/graphene heterostructure for zero twist angle, calculated on a $90\times90\times1$ $k$-point mesh. (a) Total DOS (blue) compared with the DOS summed over the atomic projections of all atoms (orange). The small deviation between the two curves reflects the fact that the atomic-orbital projection basis, being finite, does not perfectly reproduce the full plane-wave wavefunctions, so a small fraction of the spectral weight is not captured at each energy.  (b) Atom-resolved contributions to the total DOS: top (black) and bottom (red, dashed) graphene layers, W (magenta), and Se (green). The top and bottom graphene contributions are identical across the full energy range, confirming the $\sigma_{\rm h}$ symmetry of the heterostructure. Graphene states dominate the total DOS near the Fermi level, while the W and Se contributions are comparatively small (see inset). Note that the individual atomic contributions in (b) are shown on a DOS/3 
    scale for visual clarity.}
    \label{fig:pdos}
\end{figure}

Figure~\ref{fig:pdos} shows the atom-projected density of states (pDOS) of the studied heterostructure for zero twist angle, calculated on a $90\times90\times1$ $k$-point mesh. In panel (a), we compare the total density of states with the density of states summed over the atomic projections of all atoms in the cell; the two do not coincide exactly, indicating that the atomic projection does not capture the full spectral weight of the true wavefunctions (see inset of (a)). Despite this, the atomic projection allows us to resolve the individual atomic contributions, shown in panel (b). Close to the Fermi level, the graphene states dominate over the W and Se contributions (see inset of (b)), with the top and bottom carbon layers contributing equally, a direct consequence of the $\sigma_{\rm h}$ mirror plane symmetry of the heterostructure. Thus, one can model the behavior of the heterostructure at the Fermi level using only the effective model of synthetic bilayer graphene.

\section{Fitting procedure and effective model parameters of graphene(s)}
In Table~1 of the main text, we reported effective parameters for the top and bottom graphene layers within the graphene/WSe$_2$/graphene heterostructure for zero twist angle $\theta=0^{\rm o}$.
The effective Hamiltonian of synthetic bilayer graphene in the $\theta=0^{\rm o}$ case is derived under the assumption that the individual graphene/WSe$_2$ stack has the ${\bf C}_{3{\rm v}}$ symmetry. However, this is not the case for $\theta\neq0^{\rm o}$, where the common symmetry is reduced to ${\bf C}_{3}$. In this case, the Hamiltonian of proximitized graphene is equal to $H_{\rm gr}=H_0+H_{\rm I}+H_{\rm{R}}(\phi_{\rm R})$, where the only difference with the proximitized graphene compatible with the ${\bf C}_{3{\rm v}}$ symmetry is the appearance of the Rashba phase ${\phi_{\rm R}}$~\cite{David2019,Li2019b} in the Rashba Hamiltonian. The exact form of the Rashba Hamiltonian compatible with the ${\bf C}_{3}$ symmetry can be found in~\cite{MMJ+26}.

Having this in mind, here we present the effective parameters of the effective synthetic graphene bilayer created in four different graphene/WSe$_2$/graphene heterostructures, characterized by the twist angles $\theta=0^{\rm o}$, $5.2^{\rm o}$, $19.1^{\rm o}$, and $27.0^{\rm o}$, between the top/bottom graphene and WSe$_2$ monolayer. Besides that, we have analyzed the influence of the perpendicular electric field $E\in[0,1]$~V/nm in the steps of 0.2 V/nm. The effective parameters were obtained by fitting the band structure and spin expectation values of graphene's around the $K=4\pi/3a_{0}^{\rm str}(1,0)$ point, whereas the $k$-path was following the $\Gamma K M$ line that lies on $k_y=0$. The $k$-point fitting range is extended to a maximal distance of $0.0025\,$\AA$^{-1}$ from the K point, corresponding roughly to a [-25,25]\, meV energy window around the Fermi level.   

The results are gathered Tables~\ref{TAB:parameters1}-\ref{TAB:parameters4}, where we notice that
for $E=0$ top and bottom graphene have equal  $v_{\rm F}$, $\mu$, $\Delta$, $\alpha_{\rm I}^{\rm{A/B}}$, $\phi_{\rm R}$ parameters ($\phi_{\rm R}$ exists for $\theta\neq0^{\rm o}$ only), while the sign of Rashba parameters $\alpha_{\rm R}^{\rm t}$
and $\alpha_{\rm R}^{\rm b}$ is opposite ($\alpha_{\rm R}^{\rm t}=-\alpha_{\rm R}^{\rm b}$). This is the consequence of the horizontal mirror plane symmetry ${\sigma}_{\rm h}$ that connects the parameters of top and bottom graphene. The situation is changed for nonzero electric field, which breaks the ${\sigma}_{\rm h}$ symmetry and the above relation between the effective parameters of top and bottom graphene. However, our analysis reveals that the asymmetry between top and bottom graphene induced by the perpendicular electric field primarily affects the chemical potentials and leads to $\mu^{\rm b}\neq\mu^{\rm t}$. From the point view of spin physics, broken horizontal mirror plane symmetry allows for the appearance of in-plane spin expectation values. To model such a situation we additionally allow for the nonequivalence of the Rashba parameters (the relation $\alpha_{\rm R}^{\rm t}=-\alpha_{\rm R}^{\rm b}$ does not hold for $E\neq0$); in the nozero twist angle case, $\theta\neq0^{\rm o}$, besides broken $\alpha_{\rm R}^{\rm t}=-\alpha_{\rm R}^{\rm b}$ relation, nonequivalence between $\phi_{\rm R}^{\rm t}$ and $\phi_{\rm R}^{\rm b}$ is trigerred.

The fitting procedure does not automatically determine the top/bottom labeling. To fix this issue, we performed an atom-projected analysis of the calculated bands onto the top and bottom carbon layers at finite electric field $E=1$~V/nm, for each of the four twist angles studied. This analysis shows that the bands with dominant weight on the top-layer carbon orbitals lie at positive energy relative to $E_{\rm F}$, while those with dominant weight on the bottom-layer carbon orbitals lie at negative energy, confirming that the sign of the fitted $\mu^{\rm t/b}$ is correctly assigned. Since the sign of the Rashba parameter is connected to the appropriate layer, the labeling at $E=1$~V/nm is extrapolated to the $E=0$ case used in the main text, since the electric field represents only a perturbation to the band structure at zero field.

Analysis of the results in Tables~\ref{TAB:parameters1}--\ref{TAB:parameters4} reveals strong interlayer entanglement across all configurations, dominated by the spin-independent hopping parameters $t_{\rm{A}}$ and $t_{\rm{B}}$ that set the primary energy scale. The proximity-induced SOC, characterized by Rashba strengths $\alpha_{\rm R}^{\rm t/b}$ and intrinsic SOC terms $\alpha_{\rm I}^{\rm t/b}$, matches this scale even under twist or electric fields. These comparable energy scales of proximity-induced SOC and interlayer entanglement confirm the main text conclusion that the studied heterostructure forms a synthetic graphene bilayer at the Fermi level.

\begin{table*}[t]
\caption{Parameters of the effective tight-binding model of top and bottom graphene within the graphene/WSe$_2$/graphene heterostructure with zero ($\theta=0^{\circ}$) twist angle between the top/bottom graphene and WSe$_2$ monolayer. Whereas the electric field $E$ values are given in V/nm, Fermi velocity $v_{\rm F}$ in $10^6\,$m/s, all other parameters are given in meV. }\label{TAB:parameters1}
\centering
\small
\setlength{\tabcolsep}{4.6pt}
\renewcommand{\arraystretch}{1.0}
\begin{tabular}{ccccccccccccccccc}
\hline\hline
$E$&$v_{\rm F}$&$\Delta$&$\mu^{\rm t}$&$\mu^{\rm b}$&$\alpha_{\rm R}^{\rm t}$&$\alpha_{\rm R}^{\rm b}$&$\alpha_{\rm I}^{\rm A}$&$\alpha_{\rm I}^{\rm B}$&$t_{\rm{A}}$&$t_{\rm{B}}$&$\tau_{\rm{A}}$&$\tau_{\rm{B}}$\\\hline
1&   0.812&-0.671&3.137&-5.037 &-0.566& 0.534&1.183&-1.137&7.409&6.182&-0.266&-0.271\\\hline
0.8& 0.812&-0.674&2.454&-4.395 &-0.560& 0.531&1.188&-1.132&7.271&6.148&-0.264&-0.276\\\hline
0.6& 0.812&-0.685&1.120&-3.388 &-0.561& 0.540&1.199&-1.125&7.122&6.167&-0.256&-0.284\\\hline
0.4& 0.812&-0.680&1.152&-3.307 &-0.560& 0.540&1.199&-1.126&7.129&6.151&-0.255&-0.284\\\hline
0.2& 0.812&-0.411&-0.135&-2.172&-0.548& 0.536&1.173&-1.162&7.251&5.915&-0.198&-0.338\\\hline
  0& 0.813&-0.343&-1.142&-1.142&-0.538& 0.538&1.170&-1.174&7.199&5.939&-0.171&-0.358\\\hline
\end{tabular}
\end{table*} 

\begin{table*}[t]
\caption{Parameters of the effective tight-binding model of top and bottom graphene within the graphene/WSe$_2$/graphene heterostructure with nonzero ($\theta=5.2^{\circ}$) twist angle between the top/bottom graphene and WSe$_2$ monolayer. The electric field $E$ values are given in V/nm, Fermi velocity $v_{\rm F}$ in $10^6\,$m/s, Rashba angles $\phi_{\rm R}$ in degrees, while all other parameters are given in meV.}\label{TAB:parameters2}
\centering
\footnotesize
\setlength{\tabcolsep}{3pt}
\renewcommand{\arraystretch}{1.0}
\begin{tabular}{ccccccccccccccccccc}
\hline\hline
$E$&$v_{\rm F}$&$\Delta$&$\mu^{\rm t}$&$\mu^{\rm b}$&$\alpha_{\rm R}^{\rm t}$&$\alpha_{\rm R}^{\rm b}$&$\phi_{\rm R}^{\rm t}$&$\phi_{\rm R}^{\rm b}$&$\alpha_{\rm I}^{\rm A}$&$\alpha_{\rm I}^{\rm B}$&$t_{\rm{A}}$&$t_{\rm{B}}$&$\tau_{\rm{A}}$&$\tau_{\rm{B}}$\\\hline
 1.0& 0.839& 0.049& 1.917&-4.017& -0.604&0.594&-1.850 &-2.136 &1.834&-0.426&6.583&6.699&0.192&-0.867\\\hline
 0.8& 0.839& 0.044& 1.726&-3.831& -0.602&0.594&-1.815 &-2.113 &1.833&-0.428&6.569&6.679&0.197&-0.871\\\hline
 0.6& 0.839& 0.011& 0.598&-2.860& -0.583&0.581&-1.626 &-1.901 &1.820&-0.445&6.513&6.567&0.222&-0.894\\\hline
 0.4& 0.839& 0.001&-0.728&-1.447& -0.567&0.567&-0.480 &-0.635 &1.795&-0.471&6.531&6.463&0.275&-0.957\\\hline
 0.2& 0.839& 0.005& 0.361&-2.443& -0.575&0.575&-1.219 &-1.563 &1.813&-0.453&6.508&6.540&0.235&-0.904\\\hline
   0& 0.839& 0.015&-1.160&-1.160& -0.560&0.560&-19.778&-19.778&1.783&-0.483&6.543&6.454&0.296&-0.969\\\hline
\end{tabular}
\end{table*} 

\begin{table*}[t]
\caption{Parameters of the effective tight-binding model of top and bottom graphene within the graphene/WSe$_2$/graphene heterostructure with nonzero ($\theta=19.1^{\circ}$) twist angle between the top/bottom graphene and WSe$_2$ monolayer. The electric field $E$ values are given in V/nm, Fermi velocity $v_{\rm F}$ in $10^6\,$m/s, Rashba angles $\phi_{\rm R}$ in degrees, while all other parameters are given in meV.}\label{TAB:parameters3}
\centering
\footnotesize
\setlength{\tabcolsep}{3pt}
\renewcommand{\arraystretch}{1.0}
\begin{tabular}{ccccccccccccccccccc}
\hline\hline
$E$&$v_{\rm F}$&$\Delta$&$\mu^{\rm t}$&$\mu^{\rm b}$&$\alpha_{\rm R}^{\rm t}$&$\alpha_{\rm R}^{\rm b}$&$\phi_{\rm R}^{\rm t}$&$\phi_{\rm R}^{\rm b}$&$\alpha_{\rm I}^{\rm A}$&$\alpha_{\rm I}^{\rm B}$&$t_{\rm{A}}$&$t_{\rm{B}}$&$\tau_{\rm{A}}$&$\tau_{\rm{B}}$\\\hline
 1.0& 0.792&0.196& 1.433&-3.208&-0.786&0.780& 56.194 & 55.420&0.576&-0.507& 5.469 &5.169 & -0.737&-0.670\\\hline
 0.8& 0.792&0.173& 0.769&-2.727&-0.792&0.791& 56.014 & 55.402&0.567&-0.516& 5.502 &5.137 & -0.770&-0.639\\\hline
 0.6& 0.792&0.214& 0.392&-1.960&-0.802&0.795& 17.664 & 17.560&0.614&-0.475& 5.446 &5.202 & -0.761&-0.642\\\hline 
 0.4& 0.792&0.152&-1.144&-1.314&-0.808&0.806&-57.308 &-52.048&0.661&-0.459& 5.394 &5.267 & -0.745&-0.627\\\hline
 0.2& 0.792&0.171&-1.024&-1.225&-0.812&0.809&-57.778 &-51.894&0.669&-0.461& 5.408 &5.256 & -0.743&-0.620\\\hline
   0& 0.792&0.090&-1.221&-1.221&-0.802&0.802&-56.764 &-56.764&0.611&-0.474& 5.358 &5.287 & -0.841&-0.561\\\hline
\end{tabular}
\end{table*}

\begin{table*}[t]
\caption{Parameters of the effective tight-binding model of top and bottom graphene within the Gr/WSe$_2$/Gr heterostructure with nonzero ($\theta=27.0^{\circ}$) twist angle between the top/bottom graphene and WSe$_2$ monolayer. The electric field $E$ values are given in V/nm, Fermi velocity $v_{\rm F}$ in $10^6\,$m/s, Rashba angles $\phi_{\rm R}$ in degrees, while all other parameters are given in meV.}\label{TAB:parameters4}
\centering
\footnotesize
\setlength{\tabcolsep}{3pt}
\renewcommand{\arraystretch}{1.0}
\begin{tabular}{ccccccccccccccccccc}
\hline\hline
$E$&$v_{\rm F}$&$\Delta$&$\mu^{\rm t}$&$\mu^{\rm b}$&$\alpha_{\rm R}^{\rm t}$&$\alpha_{\rm R}^{\rm b}$&$\phi_{\rm R}^{\rm t}$&$\phi_{\rm R}^{\rm b}$&$\alpha_{\rm I}^{\rm A}$&$\alpha_{\rm I}^{\rm B}$&$t_{\rm{A}}$&$t_{\rm{B}}$&$\tau_{\rm{A}}$&$\tau_{\rm{B}}$\\\hline
 1.0& 0.833& -0.061& 1.671&-4.056& -0.859&0.843&-16.553&-16.020&0.273&-0.324&5.571&5.616&-0.197&-0.075\\\hline
 0.8& 0.833& -0.072& 1.390&-3.810& -0.858&0.846&-16.633&-16.003&0.270&-0.328&5.498&5.570&-0.199&-0.074\\\hline
 0.6& 0.833& -0.096& 0.084&-2.920& -0.867&0.843&-15.871&-15.923&0.234&-0.383&5.169&5.470&-0.250&-0.050\\\hline
 0.4& 0.833& -0.110&-0.202&-1.953& -0.900&0.864&-16.312&-16.112&0.278&-0.333&5.179&5.368&-0.213&-0.121\\\hline
 0.2& 0.833& -0.166&-0.818&-1.843& -0.895&0.859&-16.255&-16.158&0.299&-0.312&5.153&5.345&-0.200&-0.162\\\hline
   0& 0.833& -0.146&-1.196&-1.196& -0.914&0.914&  1.782& 1.782 &0.314&-0.305&5.214&5.223&-0.138&-0.147\\\hline
\end{tabular}
\end{table*}

\section{Charge-spin conversion coefficients dependence on the twist angle}

To further understand the implications of the synthetic graphene bilayer formation on the spin physics of the individual graphene layers, we analyze the charge-spin conversion (CSC) coefficients as a function of two independent tuning parameters: the twist angle $\theta$ between the graphene layers and the WSe$_2$ monolayer, and a perpendicular electric field $E$. We first focus on the role of the twist angle and then turn to the effect of a finite electric field.
The model parameters used in this analysis can be found in Tables~\ref{TAB:parameters1}-\ref{TAB:parameters4}.

In the main text, we introduced the conventional Rashba-Edelstein coefficient $\alpha_{\rm REE}$, which quantifies the spin accumulation $\delta S_y$ perpendicular to the applied charge current $\delta J_x$. For a nonzero twist angle $\theta$ between graphene and WSe$_2$, a finite Rashba phase $\phi_{\rm R}$ (see Sec.~2) appears, leading to the canting of the in-plane spin texture. This canting gives rise to a spin accumulation component collinear with the applied current, quantified by the unconventional Rashba-Edelstein coefficient
\begin{equation}
\alpha_{\rm UREE} = \frac{e v_{\rm F}}{\hbar}\frac{\delta S_x}{\delta J_x},
\end{equation}
where $\delta S_x$ is the current-induced nonequilibrium spin density along the $x$ axis, i.e., parallel to the applied current. Both $\alpha_{\rm REE}$ and $\alpha_{\rm UREE}$ are evaluated using the same Kubo formula in the Smrčka-Středa formulation~\cite{SS77,CB01,BM20}. 
For the results presented in this Section, the integration of $\chi_{\mathcal{O}}^{\rm sea}$ and $\chi_{\mathcal{O}}^{\rm surf}$ around the $K$ and $K'$ points was carried out on a square grid, centered around $K(K')$, having the size of $0.008$~\AA$^{-1}$ and the step $\Delta k=5\times10^{-7}$~\AA$^{-1}$, at an electronic temperature $k_{\rm B}T=0.01$~meV and quasiparticle lifetime parameter $\gamma=3.3\times10^{-3}$~meV, consistent with the values used in the main text.

\begin{figure}[t]
    \centering
\includegraphics[width=0.99\linewidth]{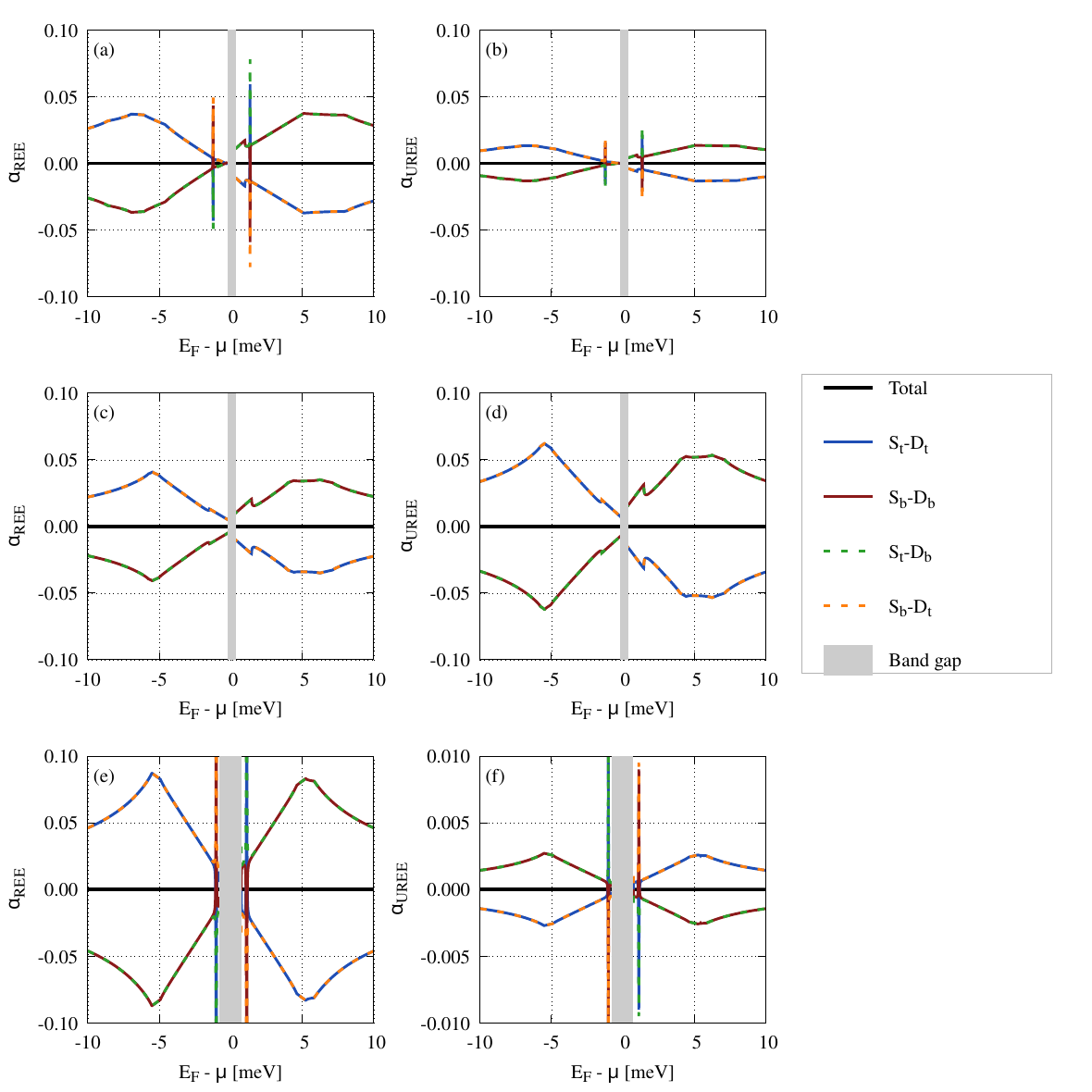}
    \caption{Charge-spin conversion coefficients for $E = 0$~V/nm, shown for three twist angles $\theta$ between the graphene layers and WSe$_2$: (a) $\alpha_{\rm REE}$ and (b) $\alpha_{\rm UREE}$ for $\theta = 5.2^\circ$; (c) $\alpha_{\rm REE}$ and (d) $\alpha_{\rm UREE}$ for $\theta = 19.1^\circ$; (e) $\alpha_{\rm REE}$ and (f) $\alpha_{\rm UREE}$ for $\theta = 27.0^\circ$. Each panel shows the total $\alpha_{\rm REE}$ and  $\alpha_{\rm UREE}$ coefficient, together with the S$_{\rm t}$-D$_{\rm t}$, S$_{\rm b}$-D$_{\rm b}$, S$_{\rm t}$-D$_{\rm b}$, and S$_{\rm b}$-D$_{\rm t}$ contributions. The chemical potential $\mu =\mu^{\rm t}=\mu^{\rm b}$ is equal to the top (bottom)-layer chemical potential. The gray rectangle represents the  gap for each twist angle, where $\alpha_{\rm REE}$ and $\alpha_{\rm UREE}$ are not properly defined.}\label{fig:(U)REE_zero}
\end{figure}

Figure~\ref{fig:(U)REE_zero} shows the doping dependence of $\alpha_{\rm REE}$ and $\alpha_{\rm UREE}$ for the three nonzero twist angles at $E=0$. 
The nonzero cross-layer signals $\alpha_{\rm REE}^{\rm S_{\rm t}-D_{\rm b}}$ and $\alpha_{\rm REE}^{\rm S_{\rm b}-D_{\rm t}}$ ($\alpha_{\rm UREE}^{\rm S_{\rm t}-D_{\rm b}}$ and $\alpha_{\rm UREE}^{\rm S_{\rm b}-D_{\rm t}}$), together with the opposite-sign lateral signals $\alpha_{\rm REE}^{\rm S_{\rm t}-D_{\rm t}}$ and $\alpha_{\rm REE}^{\rm S_{\rm b}-D_{\rm b}}$ ($\alpha_{\rm UREE}^{\rm S_{\rm t}-D_{\rm t}}$ and $\alpha_{\rm UREE}^{\rm S_{\rm b}-D_{\rm b}}$), observed for all three twist angles, additionally strengthen the conclusions drawn in the main text. We note that the sharp peaks, appearing away from the $K$ point, originate from band crossings away from the $K$ point. At these points, Fermi sea susceptibility produces peaks of the CSC coefficients, because of the vanishing energy denominator $(\epsilon_{n{\bf k}}-\epsilon_{m{\bf k}})^2 \to 0$. Their precise energy position depends on the interlayer tunneling amplitudes.

It is important to mention that, for each twist angle, $\alpha_{\rm REE}$ and $\alpha_{\rm UREE}$ have an identical functional dependence on doping, dependent on the single multiplicative factor. 
This stems from the fact that the spin operator that enters the definitions of both REE and UREE coefficients is projected onto the drain layer, where the spin accumulation is measured; the local Rashba spin texture of a given layer is characterized by its Rashba phase $\phi_{\rm R}$, and the ratio of the two coefficients for a fixed drain layer is equal to $\alpha_{\rm REE}/\alpha_{\rm UREE} = \cot{\phi_{\rm R}}$, independent of the chemical potential.
As a result of that, the S$_{\rm t}$-D$_{\rm t}$ and S$_{\rm b}$-D$_{\rm t}$ configurations (both with the drain in the top layer) have a ratio dependent on $\phi_{\rm R}^{\rm t}$, while S$_{\rm b}$-D$_{\rm b}$ and S$_{\rm t}$-D$_{\rm b}$ (drain in the bottom layer) have a ratio set by $\phi_{\rm R}^{\rm b}$. Since for $E=0$~V/nm, the horizontal mirror symmetry demands $\phi_{\rm R}^{\rm t}=\phi_{\rm R}^{\rm b}$ (Tables~\ref{TAB:parameters2}-\ref{TAB:parameters4}), all four configurations have the same ratio $\cot{\phi_{\rm R}}$, explaining why $\alpha_{\rm REE}$ and $\alpha_{\rm UREE}$ look like rescaled copies of one another in Fig.~\ref{fig:(U)REE_zero}. 

\begin{figure}[t]
    \centering
\includegraphics[width=0.99\linewidth]{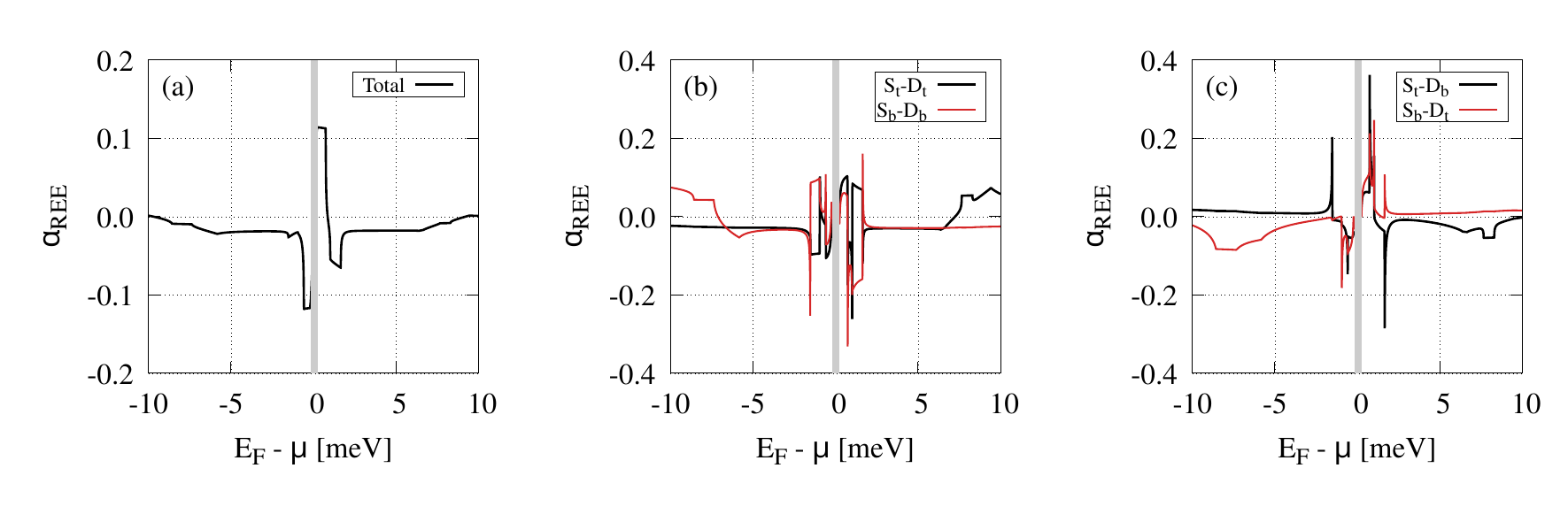}
    \caption{Charge-spin conversion coefficients for $\theta = 0^\circ$ and $E = 1$~V/nm: (a) doping dependence of the total REE coefficient $\alpha_{\rm REE}$; (b) $\alpha_{\rm REE}$ for the S$_{\rm t}$-D$_{\rm t}$ and S$_{\rm b}$-D$_{\rm b}$ contributions; (c) $\alpha_{\rm REE}$ for the S$_{\rm t}$-D$_{\rm b}$ and S$_{\rm b}$-D$_{\rm t}$ contributions. The chemical potential $\mu_{0} = (\mu^{\rm{t}} + \mu^{\rm{b}})/2$ is determined as the average of the top and bottom chemical potentials.}\label{fig:(U)REE_angle0}
\end{figure}

We now turn to the effects of the external electric field. As discussed above, an applied perpendicular electric field breaks the horizontal mirror plane $\sigma_{\rm h}$, so that the CSC coefficients of the top and bottom graphene layers are no longer connected. We start our analysis with the zero-twist-angle case (see Fig.~\ref{fig:(U)REE_angle0}), where only the REE effect is present. Opposite to the $E=0$ case, the total coefficient $\alpha_{\rm REE}$ (Fig.~\ref{fig:(U)REE_angle0}(a)) is no longer zero, due to the broken $\sigma_{\rm h}$ symmetry. This is a consequence of different top and bottom chemical potentials, $\mu^{\rm t}$ and $\mu^{\rm b}$, for nonzero field, and the Rashba parameters $\alpha_{\rm R}^{\rm t}$ and $\alpha_{\rm R}^{\rm b}$ no longer satisfy $\alpha_{\rm R}^{\rm t}=-\alpha_{\rm R}^{\rm b}$, so that the local contributions from the top and bottom graphene layers no longer cancel. We checked that a finite doping asymmetry $\mu^{\rm t}\neq\mu^{\rm b}$ alone, even when  $\alpha_{\rm R}^{\rm t}=-\alpha_{\rm R}^{\rm b}$ is preserved, is sufficient to produce a nonzero total $\alpha_{\rm REE}$, since the two layers unequally contribute to the Fermi-surface and Fermi-sea susceptibilities, despite having opposite Rashba textures. 

Next, we analyze the layer-selective contributions S$_{\rm t}$-D$_{\rm t}$ and S$_{\rm b}$-D$_{\rm b}$, shown in Fig.~\ref{fig:(U)REE_angle0}(b), and S$_{\rm t}$-D$_{\rm b}$ and S$_{\rm b}$-D$_{\rm t}$, shown in Fig.~\ref{fig:(U)REE_angle0}(c). Under the horizontal mirror symmetry $\sigma_{\rm h}$, these pairs of configurations are related by $\alpha_{\rm REE}^{\rm S_{\rm t}-D_{\rm t}}=-\alpha_{\rm REE}^{\rm S_{\rm b}-D_{\rm b}}$ and $\alpha_{\rm REE}^{\rm S_{\rm t}-D_{\rm b}}=-\alpha_{\rm REE}^{\rm S_{\rm b}-D_{\rm t}}$, as we saw in the $E=0$ case. 
Since $\mu^{\rm t}\neq\mu^{\rm b}$ breaks $\sigma_{\rm h}$ symmetry, both antisymmetric relations are no longer valid. As a consequence of that, both the local ($\alpha_{\rm REE}^{\rm S_{\rm t}-D_{\rm t}}$, $\alpha_{\rm REE}^{\rm S_{\rm b}-D_{\rm b}}$) and nonlocal ($\alpha_{\rm REE}^{\rm S_{\rm t}-D_{\rm b}}$, $\alpha_{\rm REE}^{\rm S_{\rm b}-D_{\rm t}}$) contributions acquire distinct doping dependences, due to the asymmetric electrostatic environment of the two layers. However, the most important feature of the studied system, the nonlocal charge-spin conversion is preserved, as obvious from the nonzero $\alpha_{\rm REE}^{\rm S_{\rm t}-D_{\rm b}}$ and $\alpha_{\rm REE}^{\rm S_{\rm b}-D_{\rm t}}$ signals in Fig.~\ref{fig:(U)REE_angle0}(c).

\begin{figure}[t]
    \centering
\includegraphics[width=0.99\linewidth]{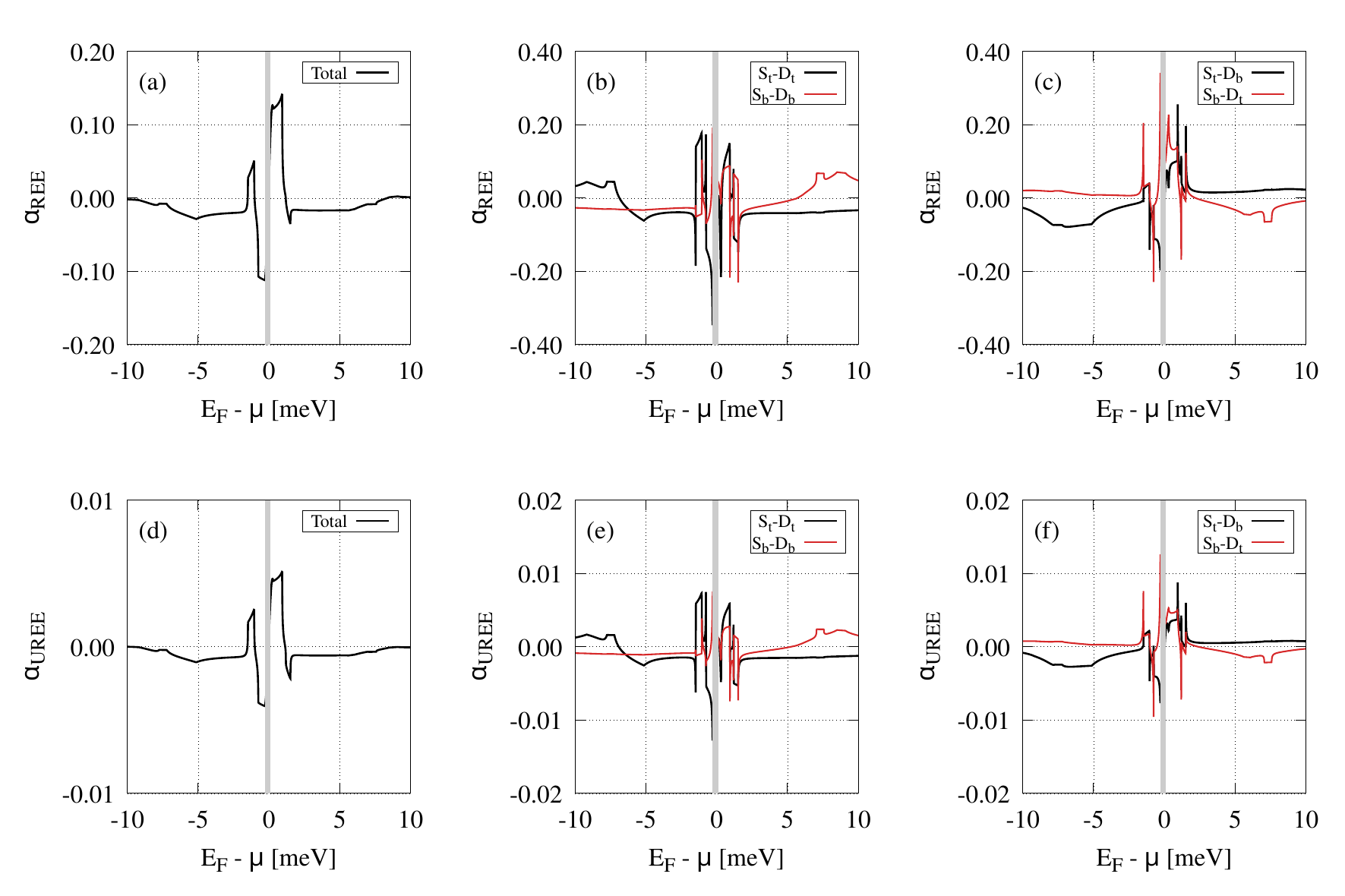}
    \caption{Charge-spin conversion coefficients for $\theta = 5.2^\circ$ and $E = 1$~V/nm: (a) doping dependence of the total REE coefficient $\alpha_{\rm REE}$; (b) $\alpha_{\rm REE}$ for the S$_{\rm t}$-D$_{\rm t}$ and S$_{\rm b}$-D$_{\rm b}$ contributions; (c) $\alpha_{\rm REE}$ for the S$_{\rm t}$-D$_{\rm b}$ and S$_{\rm b}$-D$_{\rm t}$ contributions; 
    (d) doping dependence of the total UREE coefficient $\alpha_{\rm UREE}$; (e) $\alpha_{\rm UREE}$ for the S$_{\rm t}$-D$_{\rm t}$ and S$_{\rm b}$-D$_{\rm b}$ contributions; (f) $\alpha_{\rm UREE}$ for the S$_{\rm t}$-D$_{\rm b}$ and S$_{\rm b}$-D$_{\rm t}$ contributions. The chemical potential $\mu_{0} = (\mu^{\rm{t}} + \mu^{\rm{b}})/2$ is determined as the average of the top and bottom chemical potentials.}\label{fig:(U)REE_angle52}
\end{figure}

\begin{figure}[t]
    \centering
\includegraphics[width=0.99\linewidth]{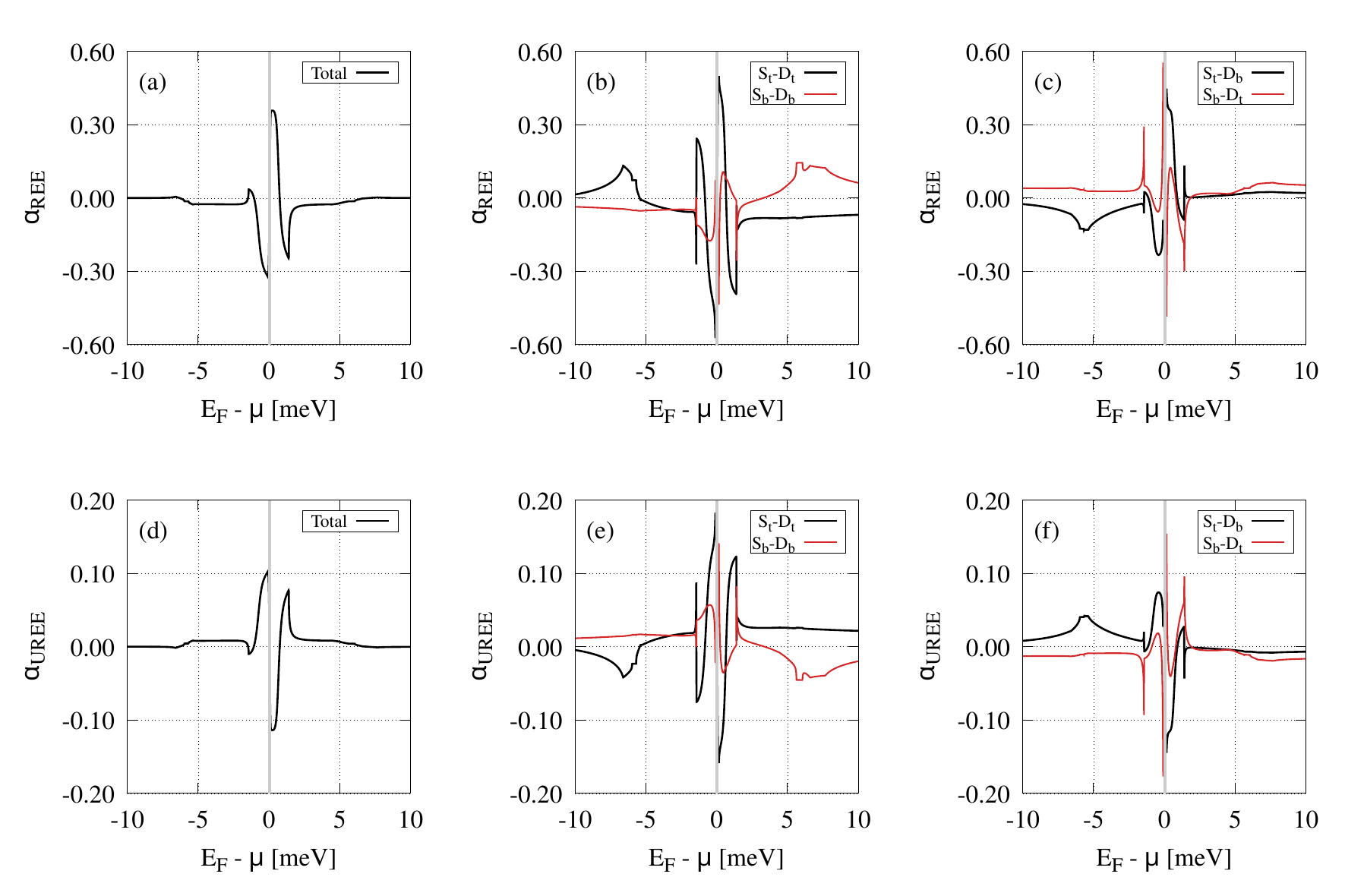}
    \caption{Charge-spin conversion coefficients for $\theta = 19.1^\circ$ and $E = 1$~V/nm: (a) doping dependence of the total REE coefficient $\alpha_{\rm REE}$; (b) $\alpha_{\rm REE}$ for the S$_{\rm t}$-D$_{\rm t}$ and S$_{\rm b}$-D$_{\rm b}$ contributions; (c) $\alpha_{\rm REE}$ for the S$_{\rm t}$-D$_{\rm b}$ and S$_{\rm b}$-D$_{\rm t}$ contributions; 
    (d) doping dependence of the total UREE coefficient $\alpha_{\rm UREE}$; (e) $\alpha_{\rm UREE}$ for the S$_{\rm t}$-D$_{\rm t}$ and S$_{\rm b}$-D$_{\rm b}$ contributions; (f) $\alpha_{\rm UREE}$ for the S$_{\rm t}$-D$_{\rm b}$ and S$_{\rm b}$-D$_{\rm t}$ contributions. The chemical potential $\mu_{0} = (\mu^{\rm{t}} + \mu^{\rm{b}})/2$ is determined as the average of the top and bottom chemical potentials.}\label{fig:(U)REE_angle191}
\end{figure}

\begin{figure}[t]
    \centering

\includegraphics[width=0.99\linewidth]{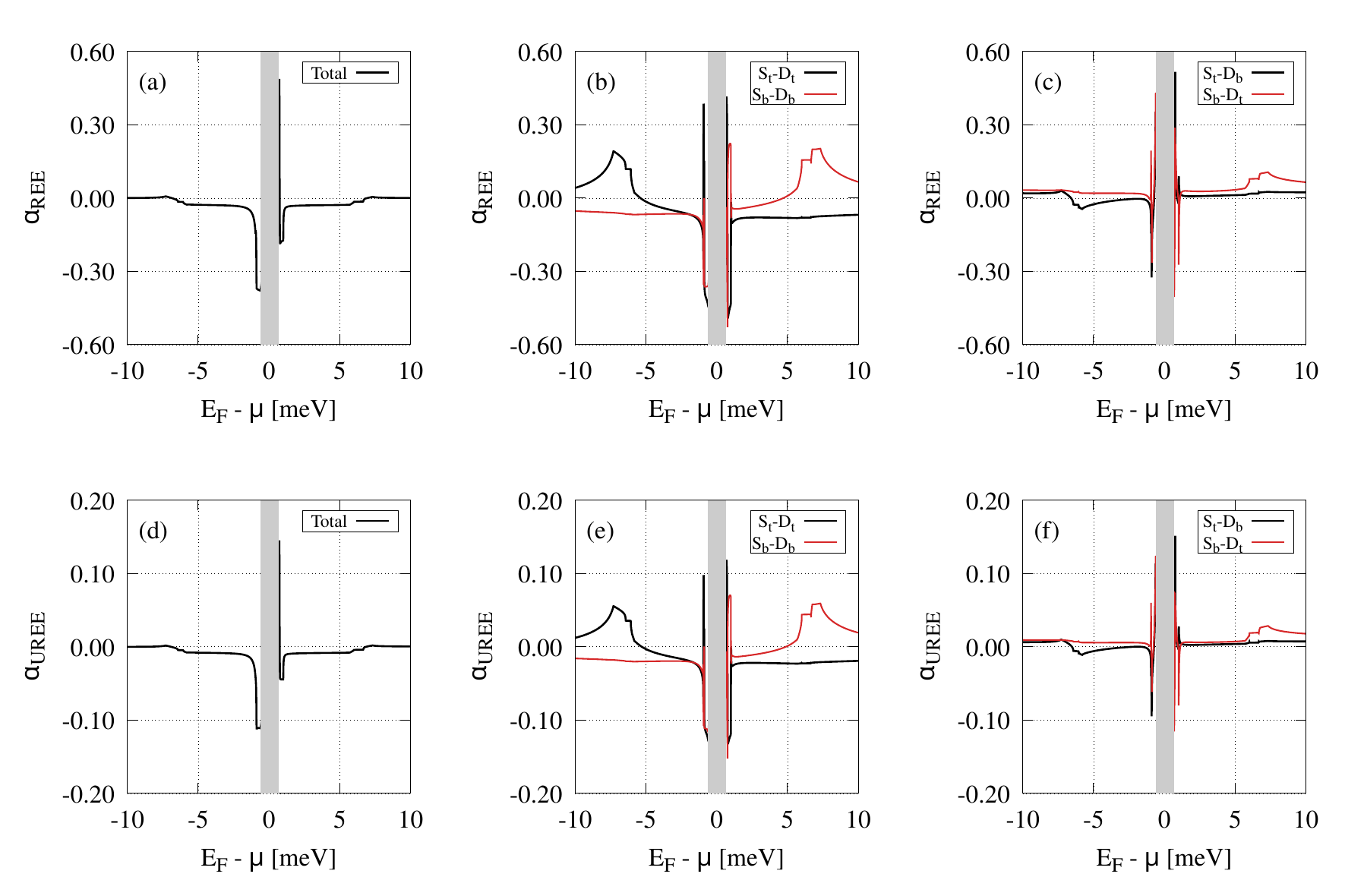}
    \caption{Charge-spin conversion coefficients for $\theta = 27.0.^\circ$ and $E = 1$~V/nm: (a) doping dependence of the total REE coefficient $\alpha_{\rm REE}$; (b) $\alpha_{\rm REE}$ for the S$_{\rm t}$-D$_{\rm t}$ and S$_{\rm b}$-D$_{\rm b}$ contributions; (c) $\alpha_{\rm REE}$ for the S$_{\rm t}$-D$_{\rm b}$ and S$_{\rm b}$-D$_{\rm t}$ contributions; 
    (d) doping dependence of the total UREE coefficient $\alpha_{\rm UREE}$; (e) $\alpha_{\rm UREE}$ for the S$_{\rm t}$-D$_{\rm t}$ and S$_{\rm b}$-D$_{\rm b}$ contributions; (f) $\alpha_{\rm UREE}$ for the S$_{\rm t}$-D$_{\rm b}$ and S$_{\rm b}$-D$_{\rm t}$ contributions. The chemical potential $\mu_{0} = (\mu^{\rm{t}} + \mu^{\rm{b}})/2$ is determined as the average of the top and bottom chemical potentials.}\label{fig:(U)REE_one}
\end{figure}

We extend our electric-field analysis to the case of finite twist angles, $\theta = 5.2^\circ$, $19.1^\circ$, and $27.0^\circ$, with the results shown in Figs.~\ref{fig:(U)REE_angle52}-\ref{fig:(U)REE_one}. In all three cases, $\alpha_{\rm UREE}$ is nonzero in addition to $\alpha_{\rm REE}$, since the finite twist angle triggers a nonzero Rashba phase $\phi_{\rm R}$, which in turn activates the unconventional Rashba-Edelstein signal. 
The applied electric field breaks the horizontal mirror-plane symmetry (thus the top-bottom equivalence). The total signal becomes nonzero, with S$_{\rm t}$-D$_{\rm t}$/S$_{\rm b}$-D$_{\rm b}$ and S$_{\rm t}$-D$_{\rm b}$/S$_{\rm b}$-D$_{\rm t}$ pairs no longer antisymmetric. Despite that, the nonlocal charge-spin conversion signals, S$_{\rm t}$-D$_{\rm b}$ and S$_{\rm b}$-D$_{\rm t}$, remain nonzero, confirming that the nonlocal spin-transport of the synthetic bilayer graphene is a robust feature of the heterostructure that persists even when both the electric field and the twist angle are simultaneously present.

\bibliographystyle{unsrt}
\bibliography{biblio}